\documentclass[runningheads]{llncs}
\usepackage[latin9]{inputenc}
\usepackage{amsmath}
\usepackage{graphicx}

\makeatletter

\providecommand{\tabularnewline}{\\}

\makeatother

\begin{document}
\title{Automatic Brain Tumor Segmentation with Scale Attention Network}
\author{Yading Yuan}
\authorrunning{Y. Yuan}
\titlerunning{Scale Attention Network for BraTS 2020}
\institute{Department of Radiation Oncology \\
 Icahn School of Medicine at Mount Sinai\\
 New York, NY, USA}
\maketitle
\begin{abstract}
Automatic segmentation of brain tumors is an essential but challenging
step for extracting quantitative imaging biomarkers for accurate tumor
detection, diagnosis, prognosis, treatment planning and assessment.
Multimodal Brain Tumor Segmentation Challenge 2020 (BraTS 2020) provides
a common platform for comparing different automatic algorithms on
multi-parametric Magnetic Resonance Imaging (mpMRI) in tasks of 1)
Brain tumor segmentation MRI scans; 2) Prediction of patient overall
survival (OS) from pre-operative MRI scans; 3) Distinction of true
tumor recurrence from treatment related effects and 4) Evaluation
of uncertainty measures in segmentation. We participate the image
segmentation challenge by developing a fully automatic segmentation
network based on encoder-decoder architecture. In order to better
integrate information across different scales, we propose a dynamic
scale attention mechanism that incorporates low-level details with
high-level semantics from feature maps at different scales. Our framework
was trained using the 369 challenge training cases provided by BraTS
2020, and achieved an average Dice Similarity Coefficient (DSC) of
$0.8828$, $0.8433$ and $0.8177$, as well as $95\%$ Hausdorff distance
(in millimeter) of $5.2176$, $17.9697$ and $13.4298$ on $166$
testing cases for whole tumor, tumor core and enhanced tumor, respectively,
which ranked itself as the 3rd place among 693 registrations in the
BraTS 2020 challenge. 
\end{abstract}

\section{Introduction}

Gliomas are the most common primary brain malignancies and quantitative
assessment of gliomas constitutes an essential step of tumor detection,
diagnosis, prognosis, treatment planning and outcome evaluation. As
the primary imaging modality for brain tumor management, multi-parametric
Magnetic Resonance Imaging (mpMRI) provides various different tissue
properties and tumor spreads. However, proper interpretation of mpMRI
images is a challenging task not only because of the large amount
of three-dimensional (3D) or four-dimensional (4D) image data generated
from mpMRI sequences, but also because of the intrinsic heterogeneity
of brain tumor. As a result, computerized analysis have been of great
demand to assist clinicians for better interpretation of mpMRI images
for brain tumor. In particular, the automatic segmentation of brain
tumor and its sub-regions is an essential step in quantitative image
analysis of mpMRI images.

The brain tumor segmentation challenge (BraTS) \cite{Menze_2015,Bakas_2017,Bakas_2018,Bakas_2017_a,Bakas_2017_b}
aims to accelerate the research and development of reliable methods
for automatic brain tumor segmentation by providing a large 3D mpMRI
dataset with ground truth annotated by multiple physicians. This year,
BraTS 2020 provides $369$ cases for model training and 125 cases
for model validation. The MRI scans were collected from 19 institutions
and acquired with different protocols, magnetic field strengths and
manufacturers. For each patient, a native T1-weighted, a post-contrast
T1-weighted, a T2-weighted and a T2 Fluid-Attenuated Inversion Recovery
(FLAIR) were provided. These images were rigidly registered, skull-stripped
and resampled to $1\times1\times1$ mm isotropic resolution with image
size of $240\times240\times155$. Three tumor subregions, including
the enhancing tumor, the peritumoral edema and the necrotic and other
non-enhancing tumor core, were manually annotated by one to four raters
following the same annotation protocol and finally approved by experienced
neuro-radiologists.

\begin{figure}
\centering\includegraphics[width=0.8\textwidth]{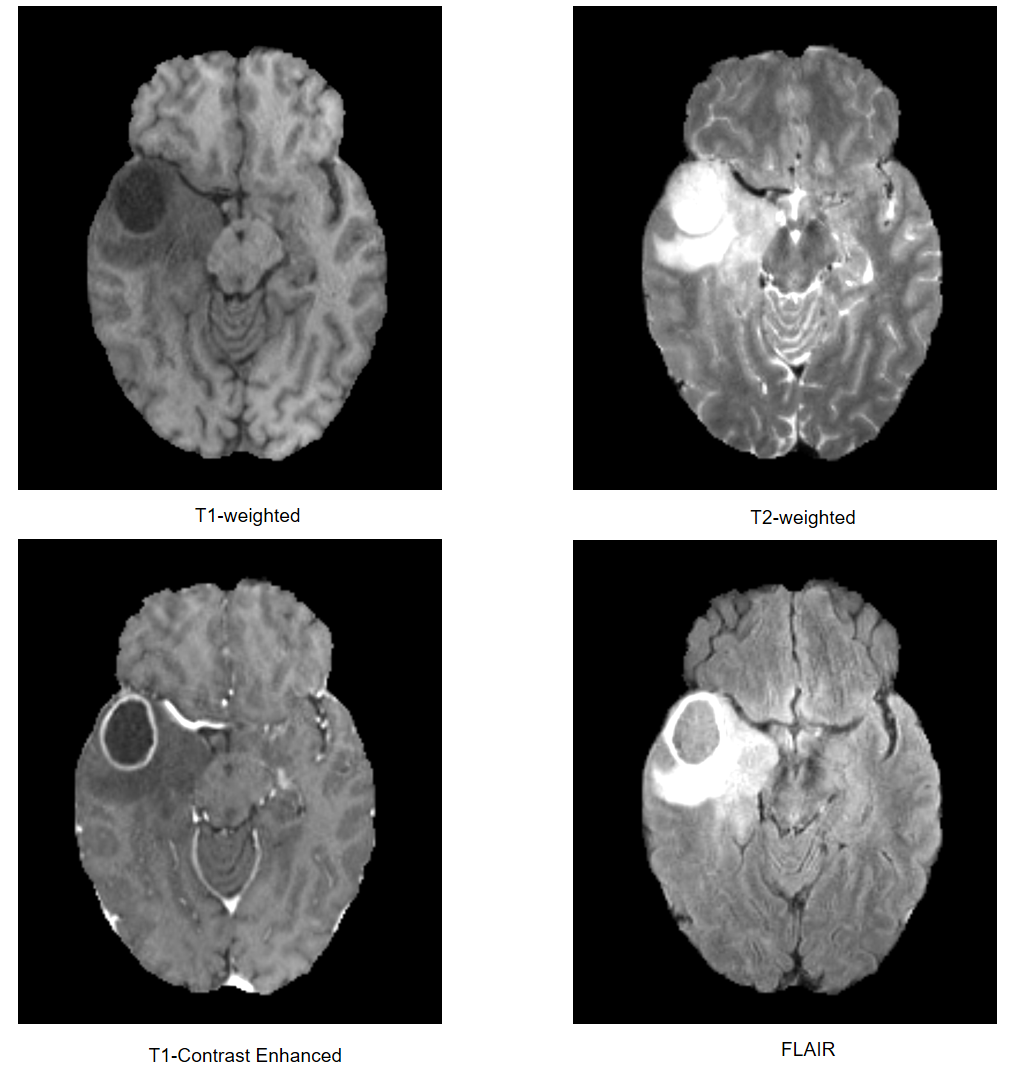}
\caption{An example of MRI modalities used in BraTS 2020 challenge}
\label{fig-mri} 
\end{figure}

\section{Related work}

With the success of convolutional neural networks (CNNs) in biomedical
image segmentation, all the top performing teams in recent BraTS challenges
exclusively built their solutions around CNNs. In BraTS 2017, Kamnitsas
et al. \cite{Kamnitsas_2018} combined three different network architectures,
namely 3D FCN \cite{Long_2015}, 3D U-Net \cite{Ronneberger_2015},
and DeepMedic \cite{Kamnitsas_2017} and trained them with different
loss functions and different normalization strategies. Wang et al.
\cite{Wang_2018} employed a FCN architecture enhanced by dilated
convolutions \cite{Chen_2018} and residual connections \cite{He_2016}.
In BraTS 2018, Myronenko \cite{Myronenko_2018} utilized an asymmetrical
U-Net with a large encoder to extract image features, and a smaller
decoder to recover the label. A variational autoencoder (VAE) branch
was added to reconstruct the input image itself in order to regularize
the shared encoder and impose additional constraints on its layers.
Isensee et al. \cite{Isensee_2019} introduced various training strategies
to improve the segmentation performance of U-Net. In BratTS 2019,
Jiang et al. \cite{Jiang_2020} proposed a two-stage cascaded U-Net,
which was trained in an end-to-end fashion, to segment the subregions
of brain tumor from coarse to fine, and Zhao et al. \cite{Zhao_2020}
investigated different kinds of training heuristics and combined them
to boost the overall performance of their segmentation model.

The success of U-Net and its variants in automatic brain tumor segmentation
is largely contributed to the skip connection design that allows high
resolution features in the encoding pathway be used as additional
inputs to the convolutional layers in the decoding pathway, and thus
recovers fine details for image segmentation. While intuitive, the
current U-Net architecture restricts the feature fusion at the same
scale when multiple scale feature maps are available in the encoding
pathway. Studies \cite{Zhou-2019,Roth-2018} have shown feature maps
in different scales usually carry distinctive information in that
low-level features represent detailed spatial information while high-level
features capture semantic information such as target position, therefore,
the full-scale information may not be fully employed with the scale-wise
feature fusion in the current U-Net architecture.

\begin{figure}
\centering\includegraphics[width=1\textwidth]{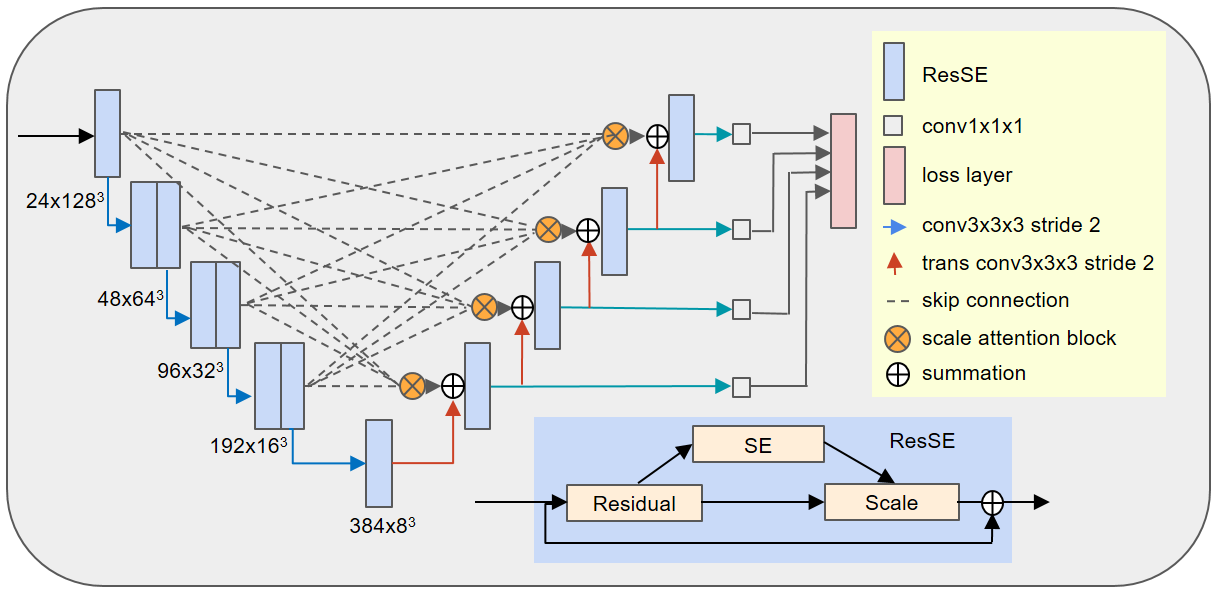}
\caption{Architecture of SA-Net. Input is a $4\times128\times128\times128$
tensor followed by one ResSE block with $24$ features. Here ResSE
stands for a squeeze-and-excitation block embedded in a residual module
\cite{Hu_2018}. By progressively halving the feature map dimension
while doubling the feature width at each scale, the endpoint of the
encoding pathway has a dimension of $384\times8\times8\times8$. The
output of the decoding pathway has three channels with the same spatial
size as the input, i.e., $3\times128\times128\times128$. }
\label{fig-sa-net} 
\end{figure}

To make full use of the multi-scale information, we propose a novel
encoder-decoder network architecture named scale attention network
(SA-Net), where we re-design the inter-connections between the encoding
and decoding pathways by replacing the scale-wise skip connections
in U-Net with full-scale skip connections. This allows SA-Net to incorporate
low-level fine details with the high-level semantic information into
a unified framework. In order to highlight the important scales, we
introduce the attention mechanism \cite{Hu_2018,Li_2019} into SA-Net
such that when the model learns, the weight on each scale for each
feature channel will be adaptively tuned to emphasize the important
scales while suppressing the less important ones. Figure \ref{fig-sa-net}
shows the overall architecture of SA-Net.

\section{Methods}

\subsection{Overall network structure}

SA-Net follows a typical encoding-decoding architecture with an asymmetrically
larger encoding pathway to learn representative features and a smaller
decoding pathway to recover the segmentation mask in the original
resolution. The outputs of encoding blocks at different scales are
merged to the scale attention blocks (SA-block) to learn and select
features with full-scale information. Due to the limit of GPU memory,
we randomly crop the input image from $240\times240\times155$ to
$128\times128\times128$, and concatenate the four MRI modalities
of each patient into a four channel tensor to yield an input to SA-Net
with the dimension of $4\times128\times128\times128$. The network
output includes three channels, each of which presents the probability
that the corresponding voxel belongs to $WT$, $TC$, and $ET$, respectively.

\subsection{Encoding pathway}

The encoding pathway is built upon ResNet \cite{He_2016} blocks,
where each block consists of two Convolution-Normalization-ReLU layers
followed by additive skip connection. We keep the batch size to 1
in our study to allocate more GPU memory resource to the depth and
width of the model, therefore, we use instance normalization \cite{Wu_2018}
that has been demonstrated with better performance than batch normalization
when batch size is small. In order to further improve the representative
capability of the model, we add a squeeze-and-excitation module \cite{Hu_2018}
into each residual block with reduction ratio $r=4$ to form a ResSE
block. The initial scale includes one ResSE block with the initial
number of features (width) of $24$. We then progressively halve the
feature map dimension while doubling the feature width using a strided
(stride=2) convolution at the first convolution layer of the first
ResSE block in the adjacent scale level. All the remaining scales
include two ResSE blocks except the endpoint of the encoding pathway,
which has a dimension of $384\times8\times8\times8$. We only use
one ResSE block in the endpoint due to its limited spatial dimension.

\subsection{Decoding pathway}

The decoding pathway follows the reverse pattern as the encoding one,
but with a single ResSE block in each spatial scale. At the beginning
of each scale, we use a transpose convolution with stride of 2 to
double the feature map dimension and reduce the feature width by 2.
The upsampled feature maps are then added to the output of SA-block.
Here we use summation instead of concatenation for information fusion
between the encoding and decoding pathways to reduce GPU memory consumption
and facilitate the information flowing. The endpoint of the decoding
pathway has the same spatial dimension as the original input tensor
and its feature width is reduced to 3 after a $1\times1\times1$ convolution
and a sigmoid function.

In order to regularize the model training and enforce the low- and
middle-level blocks to learn discriminative features, we introduce
deep supervision at each intermediate scale level of the decoding
pathway. Each deep supervision subnet employs a $1\times1\times1$
convolution for feature width reduction, followed by a trilinear upsampling
layer such that they have the same spatial dimension as the output,
then applies a sigmoid function to obtain extra dense predictions.
These deep supervision subnets are directly connected to the loss
function in order to further improve gradient flow propagation.

\subsection{Scale attention block}

\begin{figure}
\includegraphics[width=1\textwidth]{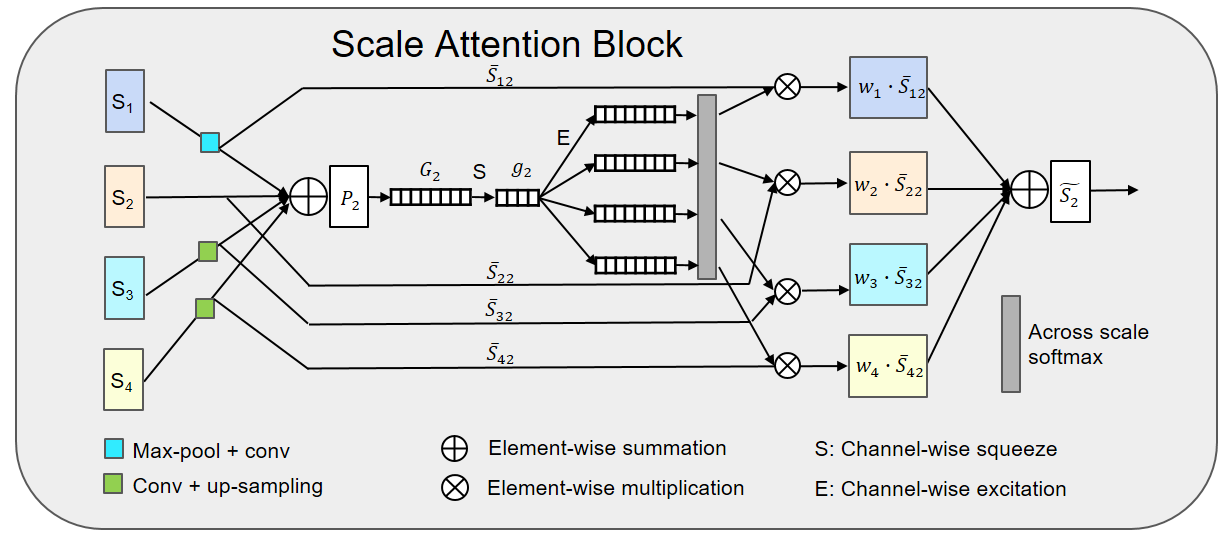} \caption{Scale attention block. Here $S1,S2,S3$ and $S4$ represent the input
feature maps at different scales from the encoding pathway. $d=2$
in this example.}
\label{fig-sa-block} 
\end{figure}

The proposed scale attention block consists of full-scale skip connections
from the encoding pathway to the decoding pathway, where each decoding
layer incorporates the output feature maps from all the encoding layers
to capture fine-grained details and coarse-grained semantics simultaneously
in full scales. As an example illustrated in Fig. \ref{fig-sa-block},
the first stage of the SA-block is to transform the input feature
maps at different scales in the encoding pathway, represented as $\{S_{e},e=1,...,N\}$
where $N$ is the number of total scales in the encoding pathway except
the last block ($N=4$ in this work), to a same dimension, i.e., $\bar{S}_{ed}=f_{ed}(S_{e})$.
Here $e$ and $d$ are the scale level at the encoding and decoding
pathways, respectively. The transform function $f_{ed}(S_{e})$ is
determined as follows. If $e<d$, $f_{ed}(S_{e})$ downsamples $S_{e}$
by $2^{(d-e)}$ times by maxpooling followed by a Conv-Norm-ReLU block;
if $e=d$, $f_{ed}(S_{e})=S_{e}$; and if $e>d$, $f_{ed}(S_{e})$
upsamples $S_{e}$ through tri-linear upsamping after a Conv-Norm-ReLU
block for channel number adjustment. After summing these transformed
feature maps as $P_{d}=\sum_e\bar{S}_{ed}$, a spatial
pooling is used to average each feature to form a information embedding
tensor $G_{d}\in R^{C_{d}}$ , where $C_{d}$ is the number of feature
channels in scale $d$. Then a $1-to-N$ Squeeze-Excitation is performed
in which the global feature embedding $G_{d}$ is squeezed to a compact
feature $g_{d}\in R^{C_{d}/r}$ by passing through a fully connected
layer with a reduction ratio of $r$, then another $N$ fully connected
layers with sigmoid function are applied for each scale excitation
to recalibrate the feature channels on that scale. Finally, the contribution
of each scale in each feature channel is normalized with a softmax
function, yielding a scale-specific weight vector for each channel
as $w_{e}\in R^{C_{d}}$, and the final output of the scale attention
block is $\widetilde{S}_{d}=\underset{e}{\sum}w_{e}\cdot\bar{S}_{ed}$.

\subsection{Implementation}

Our framework was implemented with Python using Pytorch package. As
for pre-processing, since MRI images are non-standarized, we simply
normalized each modality from each patient independently by subtracting
the mean and dividing by the standard deviation of the entire image.
The model was trained with randomly sampled patches of size $128\times128\times128$
voxels and batch size of 1. Training the entire network took 300 epochs
from scratch using Adam stochastic optimization method. The initial
learning rate was set as $0.003$, and learning rate decay and early
stopping strategies were utilized when validation loss stopped decreasing.
In particular, we kept monitoring both the validation loss ($L^{(valid)}$)
and the exponential moving average of the validation loss ($\widetilde{L}^{(valid)}$)
in each epoch. We kept the learning rate unchanged at the first $150$
epochs, but dropped the learning rate by a factor of $0.3$ when neither
$L^{(valid)}$ nor $\widetilde{L}^{(valid)}$ improved within the
last 30 epochs. The models that yielded the best $L^{(valid)}$ and
$\widetilde{L}^{(valid)}$ were recorded for model inference.

Our loss function used for model training includes two terms:

\begin{equation}
L=L_{jaccard}+L_{focal}\label{eq:loss-all}
\end{equation}
$L_{jaccard}$ is a generalized Jaccard distance loss \cite{Yuan_2017,Yuan_2017-2,Yuan_2017-3,Yuan-2019},
which we developed in our previous work for single object segmentation,
to multiple objects, and $L_{focal}$ is the voxel-wise focal loss
function that focuses more on the difficult voxels. Since the network
output has three channels corresponding to the whole tumor, tumor
core and enhanced tumor, respectively, we simply added the three loss
functions together.

In order to reduce overfitting, we randomly flipped the input volume
in left/right, superior/inferior, and anterior/posterior directions
on the fly with a probability of $0.5$ for data augmentation. We
also adjusted the contrast in each image input channel by a factor
randomly selected from {[}0.9, 1.1{]}. We used 5-fold cross validation
to evaluate the performance of our model on the training dataset,
in which a few hyper-parameters were also experimentally determined.
All the experiments were conducted on Nvidia GTX 1080 TI GPU with
11 GB memory.

During testing, we applied the sliding window around the brain region
and extracted $8$ patches with a size $128\times128\times128$ (2
windows in each dimension), and averaged the model outputs in the
overlapping regions before applying a threshold of $0.5$ to obtain
a binary mask of each tumor region.

\section{Results}

We trained SA-Net with the training set (369 cases) provided by the
BraTS 2020 challenge, and evaluated its performance on the training
set via 5-fold cross validation, as well as on the validation set,
which includes 125 cases with unknown segmentation. Table \ref{tab-cv}
shows the segmentation results in terms of Dice similarity coefficient
(DSC) for each region. As compared to the results that were obtained
from a model using the vanilla U-Net structure with scale-wise skip
connection and feature concatenation, the proposed SA-Net consistently
improved segmentation performance for each target, yielding an average
$1.47\%$ improvement.

\begin{table}
\caption{Comparison between SA-Net and U-Net segmentation results (DSC) in
5-fold cross validation using 369 training image sets. WT: whole tumor;
TC: tumor core; ET: enhanced tumor.}
\label{tab-cv}\centering %
\begin{tabular}{l|l||c|c|c|c|c|c}
\hline 
\multicolumn{1}{l}{} &  & \textbf{fold-0} & \textbf{fold-1} & \textbf{fold-2} & \textbf{fold-3} & \textbf{fold-4} & \textbf{ALL}\tabularnewline
\hline 
SA-Net & WT & 0.9256 & 0.9002 & 0.9217 & 0.9085 & 0.9195 & 0.9151\tabularnewline
 & TC & 0.8902 & 0.8713 & 0.8758 & 0.8538 & 0.8953 & 0.8773\tabularnewline
 & ET & 0.8220 & 0.7832 & 0.8198 & 0.8107 & 0.8268 & 0.8125\tabularnewline
\cline{2-8} \cline{3-8} \cline{4-8} \cline{5-8} \cline{6-8} \cline{7-8} \cline{8-8} 
 & AVG & 0.8793 & 0.8516 & 0.8724 & 0.8577 & 0.8805 & 0.8683\tabularnewline
\hline 
\hline 
U-Net & WT & 0.9218  & 0.9003  & 0.9104  & 0.9021  & 0.9113  & 0.9092\tabularnewline
 & TC & 0.8842  & 0.8735  & 0.8772  & 0.8307  & 0.8700  & 0.8671\tabularnewline
 & ET & 0.7982  & 0.7672  & 0.7922  & 0.7955  & 0.8007  & 0.7908\tabularnewline
\cline{2-8} \cline{3-8} \cline{4-8} \cline{5-8} \cline{6-8} \cline{7-8} \cline{8-8} 
 & AVG & 0.8680 & 0.8470 & 0.8599 & 0.8428 & 0.8607 & 0.8557\tabularnewline
\hline 
\end{tabular}
\end{table}

When applying the trained models on the challenge validation dataset,
a bagging-type ensemble strategy was implemented to combine the outputs
of eleven models obtained through 5-fold cross validation to further
improve the segmentation performance. In particular, our model ensemble
included the models that yielded the best $L^{(valid)}$ (validation
loss) and $\widetilde{L}^{(valid)}$(moving average of the validation
loss) respectively in each fold, plus the model that was trained with
all the $369$ cases. We uploaded our segmentation results to the
BraTS 2020 server for performance evaluation in terms of DSC, sensitivity,
specificity and Hausdorff distance for each tumor region, as shown
in Table \ref{tab-valid}.

\begin{table}
\caption{Segmentation results of SA-Net on the BraTS 2020 validation sets in
terms of Mean DSC and 95\% Hausdorff distance (mm). WT: whole tumor;
TC: tumor core; ET: enhanced tumor.}
\label{tab-valid}\centering %
\begin{tabular}{l||c|c|c|c|c|c}
\hline 
 & \multicolumn{1}{c}{} & \multicolumn{1}{c}{DSC} &  & \multicolumn{1}{c}{} & \multicolumn{1}{c}{HD95} & \tabularnewline
\hline 
 & WT  & TC  & ET  & WT  & TC  & ET\tabularnewline
\hline 
The best single model  & 0.9044  & 0.8422  & 0.7853  & 5.4912  & 8.3442  & 20.3507\tabularnewline
\hline 
Ensemble of 11 models  & 0.9108  & 0.8529  & 0.7927  & 4.0975  & 5.8879  & 18.1957\tabularnewline
\hline 
\end{tabular}
\end{table}

During the testing phase, only one submission was allowed. Table \ref{tab-test}
summarizes our final results, which ranked our method as the 3rd place
among 693 registrations in Brats 2020 challenge.

\begin{table}
\caption{Segmentation results of SA-Net on the BraTS 2020 testing sets in terms
of Mean DSC and 95\% Hausdorff distance (mm). WT: whole tumor; TC:
tumor core; ET: enhanced tumor.}
\label{tab-test}\centering %
\begin{tabular}{l||c|c|c|c|c|c}
\hline 
 & \multicolumn{1}{c}{} & \multicolumn{1}{c}{DSC} &  & \multicolumn{1}{c}{} & \multicolumn{1}{c}{HD95} & \tabularnewline
\hline 
 & WT  & TC  & ET  & WT  & TC  & ET\tabularnewline
\hline 
Ensemble of 11 models  & 0.8828  & 0.8433  & 0.8177  & 5.2176  & 17.9697  & 13.4298\tabularnewline
\hline 
\end{tabular}
\end{table}

\section{Summary}

In this work, we presented a fully automated segmentation model for
brain tumor segmentation from multimodality 3D MRI images. Our SA-Net
replaces the long-range skip connections between the same scale in
the vanilla U-Net with full-scale skip connections in order to make
maximum use of feature maps in full scales for accurate segmentation.
Attention mechanism is introduced to adaptively adjust the weights
of each scale feature to emphasize the important scales while suppressing
the less important ones. As compared to the vanilla U-Net structure
with scale-wise skip connection and feature concatenation, the proposed
scale attention block not only improved the segmentation performance
by $1.47\%$, but also reduced the number of trainable parameters
from $17.8$M (U-Net) to $16.5$M (SA-Net), which allowed it to achieve
a top performance with limited GPU resource in this challenge.

\section*{Acknowledgment}

This work is partially supported by a research grant from Varian Medical
Systems (Palo Alto, CA, USA) and grant UL1TR001433 from the National
Center for Advancing Translational Sciences, National Institutes of
Health, USA.

\end{document}